\documentclass[showpacs,aps,prd,nofootinbib,floatfix,amsmath,amssymb]{revtex4}
\usepackage{graphicx}% Include figure files
\usepackage{dcolumn}% Align table columns on decimal point
\usepackage{bm}% bold math
\usepackage{subfigure}
\usepackage{color}
\begin{document}

\title{Inflationary observables in F(R) gravity }

\author{Enrique D\'iaz, Oscar Meza-Aldama}

\affiliation{Sibatel Communications \\ 303 W Lincoln Ave No.140, Anaheim, CA 92805}%
\date{\today}% It is always \today, today,

\begin{abstract}
We present phenomenological signatures for a modified gravity model f(R), constructed with linear, quadratic, cubic and quartic terms. The obtained signatures satisfy current phenomenological bounds reported by PLANCK and BICEP3. Furthermore, two of the model solutions $\sigma_1$ and $\sigma_2$ seem to favor a much lower value for the tensor-to-scalar ratio $0.0005<r_{\sigma_1}<0.0015$ and $r_{\sigma_2}<0.00015$ than the current reported experimental bounds. The results we obtained are quantitatively similar to those presented in previous studies for $R^3$ models.
\end{abstract}


\maketitle

%-==-=-=-=-=-=-=-=-=-=-=-=-=-=-=-=-=-=-=-=-=-=-=-=-=-=-=-=-=- 
\section{Introduction}
%-==-=-=-=-=-=-=-=-=-=-=-=-=-=-=-=-=-=-=-=-=-=-=-=-=-=-=-=-=-

Inflationary cosmology ~\cite{a,b} offers an explanation for the observed flatness, homogeneity, and isotropy of the universe. The way in which one may construct an inflationary model can be accomplished in a myriad of ways, one could add higher-order terms to the inflationary potential $V(\phi)$ ~\cite{c}, some of which might be coupled to the Higgs model, others might have an origin from SUGRA ~\cite{d} or string theory \cite{Odintsov:2019mlf}. In this work, we explore an $f(R)$ theory ~\cite{felice} in which we added higher-order Ricci scalar terms as opposed to just having the linear R term given in Einstein gravity.

Several $f(R)$ models, such as the Starobinsky model $f(R)=R+\alpha R^2$, have been extensively studied and are known to produce predictions consistent with the latest observational data from the Planck satellite, particularly the spectral index and tensor-to-scalar ratio r \cite{b}, \cite{f}. These models can also be recast in the Einstein frame, where they are equivalent to scalar-tensor theories, further elucidating their connection to traditional scalar field inflationary scenarios \cite{faraoni}. In recent years, higher order polynomial functions have been put forward and their implications uncovered \cite{koreans}, \cite{indios}. More general expressions for $f(R)$, including exponential terms and other special functions, are found, for example, in \cite{Odintsov:2017hbk}, \cite{Elizalde:2017mrn}, \cite{Myrzakulov:2014hca}, \cite{Odintsov:2022rok}, and \cite{Odintsov:2021wjz}.
. For reviews on inflation in modified gravity theories, see \cite{ellis}, \cite{Nojiri:2010wj}, \cite{Odintsov:2023weg}, and \cite{Nojiri:2017ncd}. An alternative approach also exists in which, instead of proposing $f$ and from it deducing the inflationary potential $V$, one can instead propose $V$ and from it derive $f$; some examples of this approach can be seen in \cite{Sebastiani:2013eqa}.

The observational measurements of the spectral index of curvature perturbations reported by PLANCK $n_s = 0.9649 \pm 0.0042$, as well as the upper bound on the tensor-to-scalar ratio $r < 0.072$ ~\cite{f}, have been instrumental in the process of building new inflationary models. Continuing with this process, the BICEP/Keck2021  upper bound on $r < 0.036$ at $95\%$ C.L. ~\cite{e}, offers a great opportunity to see what implications this has on new models. This is why we set about and looked at a modified gravity $f(R)$ model. This formalism offers one of the simplest ways upon which new inflationary models can be constructed.

In this work, we study some of the implications of a quartic polynomial function $f(R)$, namely $f(R)=R+\frac{1}{2}\alpha R^2 + \frac{1}{2} \beta R^3 + \frac{1}{4} \gamma R^4$. The paper is structured as follows: In the next section we present a brief summary of generic inflationary $f(R)$ models, how to transform them into a regular Einstein frame theory and how to perform the calculation of the main inflationary observables in that frame. In section 3, we give the specific expressions for our quartic model. In section 4 we obtain the slow roll parameters and from them we calculate the spectral index and the tensor-to-scalar ratio, which are then plotted and compared to current experimental bounds; the bounds on the power spectrum are also used to obtain a specific central value for one of the parameters of our model. Section 5 summarizes our conclusions. Finally, in the appendix we give the acceptable solutions for the Einstein-frame field in terms of the number of e-folds.

%-==-=-=-=-=-=-=-=-=-=-=-=-=-=-=-=-=-=-=-=-=-=-=-=-=-=-=-=-=- 
\section{$f(R)$ inflationary models}
%-==-=-=-=-=-=-=-=-=-=-=-=-=-=-=-=-=-=-=-=-=-=-=-=-=-=-=-=-=- 

Inflationary models that are constructed in an $f(R)$ formalism are created by extending the Einstein Hilbert action for general relativity which is (in natural units):
\begin{equation}
    S = \frac{1}{2} \int d^4x \sqrt{-g} R.
\end{equation}
by replacing the R term with a function $f(R)$ 
\begin{equation}
    S = \frac{1}{2} \int d^4x \sqrt{-g} f(R), \label{f_R_action}
\end{equation}
where $f(R)$ is an arbitrary function of the Ricci scalar.
Usually, for calculation purposes, we introduce an auxiliary field $\phi$ and rewrite the action as \cite{koreans}:
\begin{equation}
    S = \frac{1}{2} \int d^4x \sqrt{-g} \left[ f(\phi) + f^\prime(\phi) ( R - \phi )\right].
\end{equation}
Substituting the equation of motion for $\phi$ we recover the original action \eqref{f_R_action}.
We then make the conformal transformation
\begin{equation}
    g_{\mu\nu} \rightarrow \frac{1}{f^\prime(\phi)} g_{\mu\nu},
\end{equation}
and we get the action in the Einstein frame with a minimally coupled scalar field $s$:
\begin{equation}
    S = \frac{1}{2} \int d^4x \sqrt{-g} \left[ \frac{1}{2} R - \frac{1}{2} g^{\mu\nu} \partial_\mu s \partial_\nu s - V_E(s) \right],
\end{equation}
where
\begin{gather}
    s(\phi) = \sqrt{\frac{3}{2}} \ln f^\prime (\phi), \label{s_phi} \\
    V_E(s) = \left. \frac{ \phi f^\prime (\phi) - f(\phi) }{ 2 {f^\prime}^2 (\phi) } \right|_{\phi=\phi(s)} \label{V_E_s}.
\end{gather}
These transformations are performed because calculations of the inflationary observables are usually performed in the Einstein frame. Equation \eqref{s_phi} is supposed to be inverted to get $\phi$ in terms of $s$ and use this expression to evaluate the right-hand side of equation \eqref{V_E_s}. So, if we propose a particular $f$, we must solve \eqref{s_phi} for $\phi$ in terms of $s$ and then substitute into \eqref{V_E_s} to get the potential. 

Important cosmological quantities can be calculated via the potential. For example, in the slow roll approximation, the slow roll parameters are defined as:
\begin{gather}
    \epsilon \equiv \frac{1}{2} \left( \frac{V^\prime}{V} \right)^2, \label{sr_epsilon} \\
    \eta \equiv \frac{V^{\prime\prime}}{V}. \label{sr_eta}
\end{gather}
In our case, they are functions of $s$. The end of inflation is the point $s_e$ at which $\epsilon=1$. The number of e-folds from the beginning of inflation $s$ until the end $s_e$ is defined as
\begin{equation}
    N = \int_{s_e}^{s} \frac{V}{V^\prime} ds. \label{N_Vprime_V}
\end{equation}
Integrating, we can in principle solve for $s$, i.e., $s=s(N)$. Then substitute in (8) and (9) to get $\epsilon=\epsilon(N)$ and $\eta=\eta(N)$.
Two important inflationary observables are the spectral index $n_s$ and the tensor-to-scalar ratio $r$. In the slow-roll approximation, they are given by:
\begin{gather}
    n_s \equiv 1 - 2 \epsilon_\ast - \eta_\ast, \label{n_s_def} \\
    r \equiv 16 \epsilon_\ast; \label{r_def}
\end{gather}
where the asterisk denotes evaluation at horizon exit, which is usually taken as $N_\ast \approx 60$; generally, one requires a number of 60 e-folds to solve the flatness problem ~\cite{g}.

%-==-=-=-=-=-=-=-=-=-=-=-=-=-=-=-=-=-=-=-=-=-=-=-=-=-=-=-=-=- 
\section{Quartic $f(R)$ scenario}
%-==-=-=-=-=-=-=-=-=-=-=-=-=-=-=-=-=-=-=-=-=-=-=-=-=-=-=-=-=-

The slow roll parameters have a dependency on the potential that is related to $f(R)$ via \eqref{V_E_s}.

Proposing a fourth order $f$:
\begin{equation}
    f(R) = R + \frac{1}{2}  \alpha R^2 + \frac{1}{3} \beta R^3 + \frac{1}{4} \gamma R^4,
\end{equation}
solving \eqref{s_phi} we obtain,
\begin{equation}
    e^{\sqrt{\frac{2}{3}} s} = 1 + \alpha \phi + \beta\phi^2 + \gamma\phi^3 ,
\end{equation}
which gives us $\phi$ in terms of $s$, then substituting into \eqref{V_E_s} to get $V(s)$, calculate the first derivative $V^\prime$, substitute into \eqref{N_Vprime_V}, perform the integral and solve for $s$ in terms of $N$, substitute into \eqref{r_def} and \eqref{n_s_def} and evaluate $N \approx 60$ to get $n_s$ and $r$ in terms of $\alpha$, $\beta$, $\gamma$. Following \cite{koreans}, we define
\begin{equation}
    \sigma \equiv e^{\sqrt{\frac{2}{3}} s}.
\end{equation}
The expression for $\phi$ in terms of $\sigma$ in this particular model is
\begin{equation}
\phi=\frac{(\sigma-1)}{\alpha-x}\left( \frac{\sigma-1}{\alpha} \right)^2-y\left( \frac{\sigma-1}{\alpha}\right)^3.
\end{equation}
where $x=\frac{\beta}{\alpha}$ and $y=\frac{\gamma}{\alpha}$.
This expression can be substituted into the potential and then used to calculate the inflationary observables.

%-==-=-=-=-=-=-=-=-=-=-=-=-=-=-=-=-=-=-=-=-=-=-=-=-=-=-=-=-=-
\section{Inflationary Observables}
\begin{figure}[htbp]
\centering
\includegraphics[width=.4\textwidth]{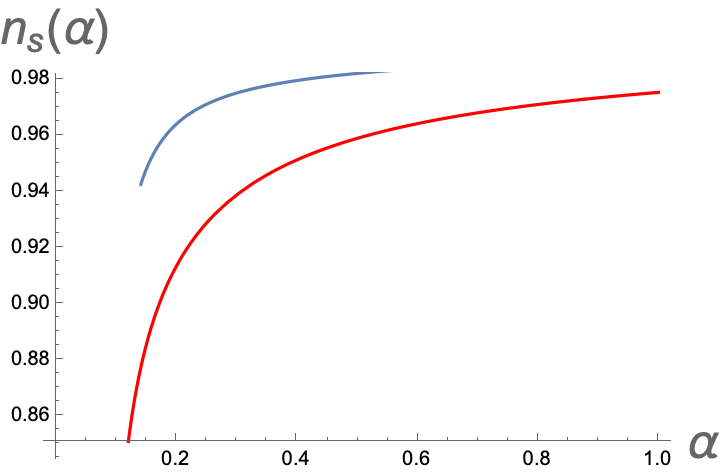}
\qquad
\includegraphics[width=.4\textwidth]{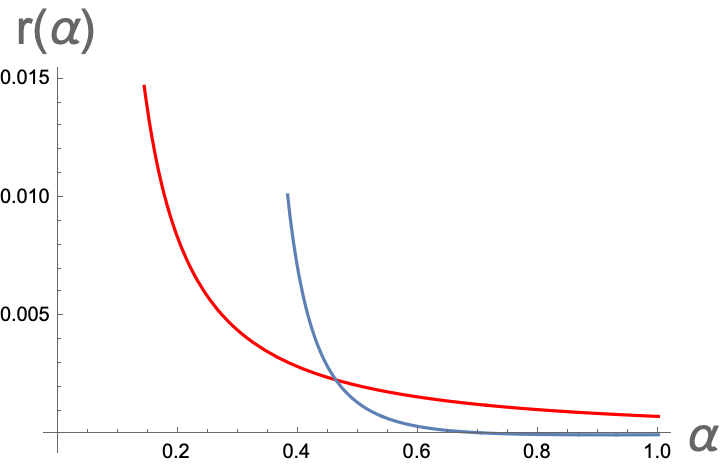}
\caption{The blue line corresponds to the solution $\sigma_1$ and the red line to the solution $\sigma _2$, we explored different values of $0<\alpha\leq 1$ with a fixed value of the number of e-folds $N=60$.\label{fig:ns_r}}
\end{figure}

\begin{figure}[htbp]
\centering
\includegraphics[width=.4\textwidth]{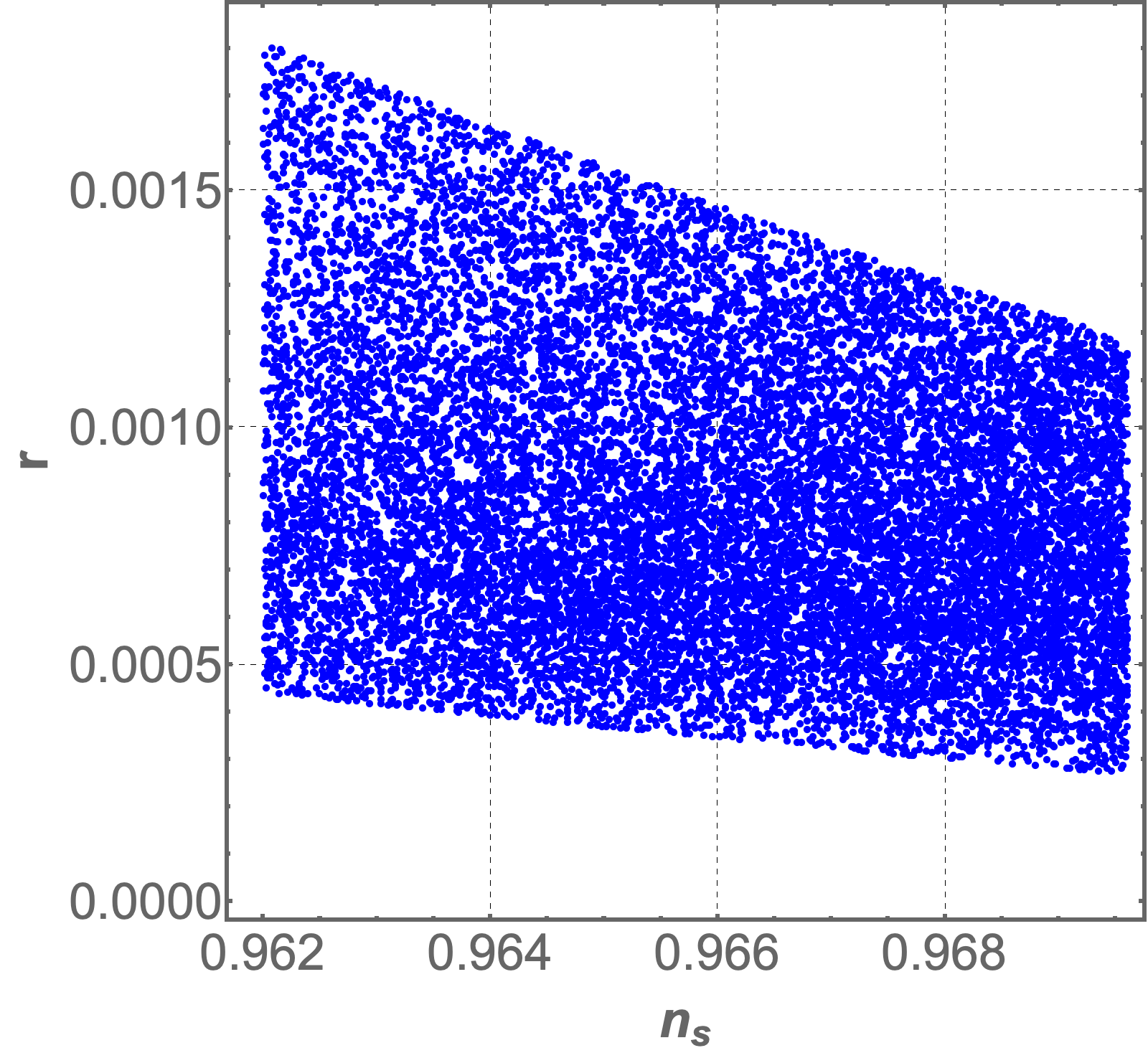}
\includegraphics[width=.4\textwidth]{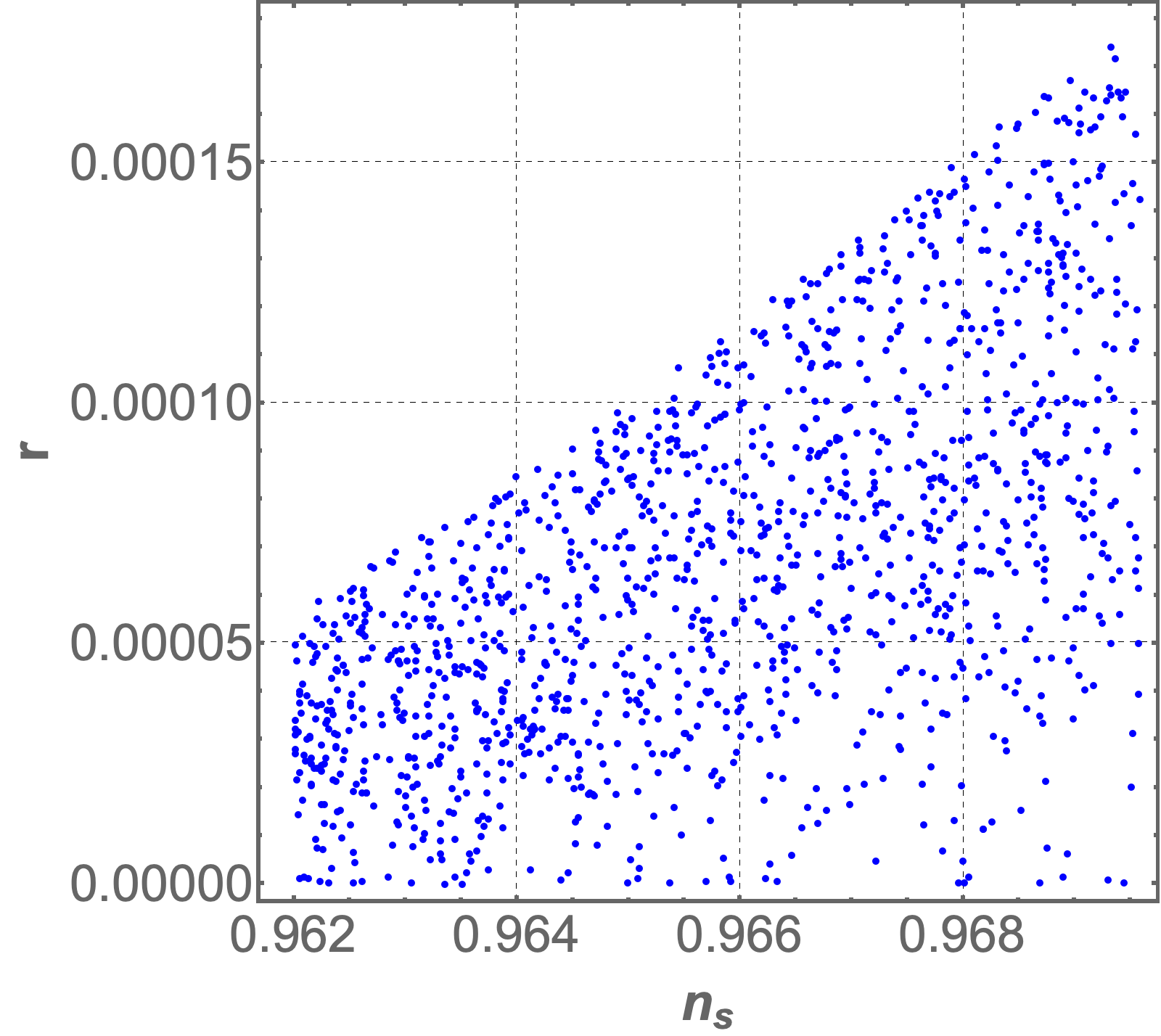}
\caption{Fits corresponding to solution $\sigma_1$ (left) and $\sigma_2$ (right), which satisfy current bounds for the spectral index $n_s$ and tensor-to-scalar ratio $r$ for a fixed value of the number of e-folds $N=60$.\label{fig:scatter_ns_r}}
\qquad
\end{figure}

%-==-=-=-=-=-=-=-=-=-=-=-=-=-=-=-=-=-=-=-=-=-=-=-=-=-=-=-=-=-
The first two slow roll parameters $\epsilon$ and $\eta$ calculated using the slow roll approximation as described in the introduction are given by:

\begin{eqnarray}
    \epsilon & = & \frac{4(-6\alpha^3+3\gamma(\sigma+1)(\sigma-1)^2+2\alpha\beta(\sigma^2+\sigma-2))^2}{3(\sigma-1)^2(-6\alpha^3+4\alpha\beta(\sigma-1)+3\gamma(\sigma-1)^2)^2}, \\
    \eta & = & \frac{8(3\alpha^3(\sigma-2)+3\gamma(\sigma-1)^2(1+\sigma+\sigma^2)+\alpha\beta(\sigma^3+3\sigma-4))}{3(-6\alpha^3 + 4\alpha\beta(\sigma-1)+3\gamma(\sigma-1)^2)(\sigma-1)^2}.
\end{eqnarray}
From these we obtain the observables $r$ and $n_s$ as given from equations \eqref{n_s_def} and \eqref{r_def}. The plots are shown in figure \ref{fig:ns_r}.
We also present a scatter plot of allowed values for $r$ and $n_s$ in figure \ref{fig:scatter_ns_r}. This plot was generated using the package presented in \cite{marco}.

The number of e-folds is
\begin{equation}
    N \equiv \int_{s_e}^s \frac{1}{\sqrt{2\epsilon}} ds;
\end{equation}
upon changing variables from $s$ to $\sigma$ 
\begin{equation}
    \sigma = e^{\sqrt{\frac{2}{3}} s},
\end{equation}
we get
\begin{equation}
    N = \sqrt{\frac{3}{2}} \int_{\sigma_e}^\sigma \frac{V(\sigma)}{\sigma V^\prime(\sigma)} d\sigma,
\end{equation}
and assuming $\sigma \gg \sigma_e$, we obtain the expression
\begin{equation}
    N = \sqrt{\frac{3}{2}}\left ( \frac{(-36y+24\alpha(x+3\alpha))}{48\sqrt{6}\alpha^2}\sigma + \frac{(54y-24x\alpha)}{48\sqrt{6}\alpha^2}\sigma^2 +\frac{(-36y+8x\alpha)}{48\sqrt{6}\alpha^2}\sigma^3 +\frac{\sqrt{\frac{3}{2}}y}{16\alpha^2}\sigma^4\right), \label{N_quartic_sigma}
\end{equation}
Throughout this paper we use the values $x=2\times 10^{-8}$ and $y=10^{-8}$.
Since \eqref{N_quartic_sigma} is a quartic polynomial, we can invert it to get $\sigma$ in terms of $N$, which of course gives four solutions $\sigma_i$, $i=1,2,3,4$ (since $N$ is a quartic polynomial on $\sigma$). Two of the solutions did not satisfy current bounds for the measured slow roll parameters, hence we only work with two of the roots which are presented in the appendix.

The expression for $\sigma$ in terms of $N$ can then be substituted into the expressions for the inflationary observables \eqref{sr_eta} and \eqref{sr_epsilon}, giving $\epsilon = \epsilon(N)$  and $\eta = \eta(N)$ . Evaluating at horizon exit, $N = 60$, we get the plots shown in figure \ref{fig:ns_r}.

On the other hand, for the power spectrum, using COBE normalization
\begin{equation}
    \frac{V}{\epsilon}N = 0.027^4,
\end{equation}
evaluating at the horizon exit and using the same values of $x$ and $y$ as before, we get a concrete value for our parameter
\begin{equation}
    \alpha = 1.187 \times 10^9.
\end{equation}
This is around the same order of magnitude as the value presented in \cite{{koreans}}.

%-==-=-=-=-=-=-=-=-=-=-=-=-=-=-=-=-=-=-=-=-=-=-=-=-=-=-=-=-=-
\section{Conclusion}
Using a quartic polynomial function in an $f(R)$ gravity model, we have obtained the cosmological observables for the corresponding inflationary potential in the Einstein frame. We successfully obtained three main cosmological observables: the power spectrum and spectral index of scalar perturbations, and the tensor-to-scalar ratio. We used them to get constraints on the parameter(s) of our model and showed that within those regions the model satisfies the most up-to-date experimental bounds. The allowed parameter space that is found falls well within the reach of the expected improved measurements expected at the LiteBIRD experiment  of $\delta r < 0.001$ ~\cite{Fuskeland}.

%-==-=-=-=-=-=-=-=-=-=-=-=-=-=-=-=-=-=-=-=-=-=-=-=-=-=-=-=-=-

\section{Acknowledgments}
We thank Gabriel Jacobo for his contributions to the calculations of inflationary observables.

%-==-=-=-=-=-=-=-=-=-=-=-=-=-=-=-=-=-=-=-=-=-=-=-=-=-=-=-=-=-

\section{Appendix A: Solution $\sigma_1$}
The two solutions for $\sigma$ in terms of $N$ are given by:
\begin{equation}
\begin{split}
\sigma_{1,2} &= 1 - \frac{2 x\alpha}{9 y}-\frac{1}{2}\sqrt{\frac{2^{\frac{10}{3}}}{3\left(\frac{y}{\alpha^2}\right)^{\frac{1}{3}}}-\frac{2^{\frac{16}{3}}N}{9\left(\frac{y}{\alpha^2}\right)^{\frac{1}{3}}}-\frac{2^{\frac{1}{3}} y}{3\alpha^2\left(\frac{y}{\alpha^2}\right)^{\frac{1}{3}}}+\frac{2^{\frac{10}{3}}x}{27\alpha \left(\frac{y}{\alpha^2}\right)^{\frac{1}{3}}}-\frac{2^{\frac{13}{3}}\alpha x}{27 y\left(\frac{y}{\alpha^2}\right)^{\frac{1}{3}}}+\frac{2^{\frac{5}{3}}(\frac{y}{\alpha^2})^{\frac{1}{3}}\alpha^2}{y}} \\
& \pm \frac{1}{2}\sqrt{\left ( -\frac{2^{\frac{10}{3}}}{3(\frac{y}{\alpha^2})^{\frac{1}{3}}} + \frac{2^{\frac{97}{3}}N}{9(\frac{y}{\alpha^2})^\frac{1}{3}} + \frac{2^{\frac{1}{3}}y}{3\alpha^2\left(\frac{y}{\alpha^2}\right)^{\frac{1}{3}}} -\frac{2^{\frac{10}{3}}x}{27\alpha \left(\frac{y}{\alpha^2} \right)^{\frac{1}{3}}} + \frac{2^{\frac{49}{3}}\alpha x}{27 y \left( \frac{y}{\alpha^2} \right)^{\frac{1}{3}}} -\frac{2^{\frac{5}{3}} \alpha^2 \left( \frac{y}{\alpha^2}\right)}{y} +m
  \right) },
\end{split}
\end{equation}

\begin{equation}
m=\frac{16\alpha^2}{y\sqrt{\frac{2^{\frac{10}{3}}}{3\left(\frac{y}{\alpha^2}\right)^{\frac{1}{3}}}-\frac{2^{\frac{97}{3}}N}{9\left( \frac{y}{\alpha^2} \right)^\frac{1}{3}} -\frac{2^{\frac{1}{3}}y}{3\alpha^2 \left( \frac{y}{\alpha^2}\right)}+\frac{2^{\frac{10}{3}}x}{27\alpha\left( \frac{y}{\alpha^2}\right)^{\frac{1}{3}}}-\frac{2^{\frac{13}{3}}\alpha x}{27y\left( \frac{y}{\alpha^2}\right)^{\frac{1}{3}}}+\frac{2^{\frac{5}{3}}\alpha^2\left( \frac{y}{\alpha^2}\right)}{y}}}
\end{equation}

where for one solution we must take the plus sign and for the other one the we must take the minus sign. For this expression we have ignored higher order terms in $x$ and $y$.

%-==-=-=-=-=-=-=-=-=-=-=-=-=-=-=-=-=-=-=-=-=-=-=-=-=-=-=-=-=-

\end{document}